\documentclass[12pt,onecolumn]{article}

\usepackage{amsmath}
\usepackage{amssymb}
\usepackage[a4paper]{geometry}
\usepackage{hyperref}
\usepackage{authblk}            
\usepackage{placeins}           
\usepackage{multirow}                   
\usepackage{booktabs}
\usepackage[T1]{fontenc}
\usepackage{sectsty}
\allsectionsfont{\sffamily}

\usepackage[]{graphicx}
\usepackage[font={sf,small},format=plain,labelfont=bf]{caption}
\usepackage{subcaption}

\usepackage[
  natbib=true,
  backend=bibtex,
  maxcitenames=2,
  style=authoryear
]{biblatex}
\usepackage{orcidlink}

\title{{\sffamily Amortized Phylodynamic Inference with Neural Bayes Estimators and Recursive Neural Networks}}
\date{}
\author[1]{Alexander E. Zarebski\,\orcidlink{0000-0003-1824-7653}\thanks{Corresponding author: \texttt{aezarebski@gmail.com}}}
\author[2,3]{Thomas Williams\,\orcidlink{0000-0002-7417-5925}}
\author[4,5]{Louis du Plessis\,\orcidlink{0000-0003-0352-6289}}
\affil[1]{MRC Biostatistics Unit, University of Cambridge, United Kingdom}
\affil[2]{ARC Centre of Excellence for the Mathematical Analysis of Cellular Systems, The University of Melbourne, Australia}
\affil[3]{School of Mathematics and Statistics, The University of Melbourne, Australia}
\affil[4]{Department of Biosystems Science and Engineering, ETH Z\"{u}rich, Switzerland}
\affil[5]{Swiss Institute of Bioinformatics (SIB), Switzerland}

\newcommand{\reff}[0]{\mathcal{R}}
\newcommand{\refft}[1]{\mathcal{R}(#1)}
\newcommand{\numPrevX}[0]{N_{\text{p}}}
\newcommand{\numPrev}[1]{N_{\text{p}}(#1)}
\newcommand{\numCumX}[0]{N_{\text{c}}}
\newcommand{\numCum}[1]{N_{\text{c}}(#1)}
\newcommand{\tree}[0]{\mathcal{T}}
\newcommand{\treeSpace}[0]{\mathbb{T}}
\newcommand{\stoppingTime}[0]{T_{\text{stop}}}
\newcommand{\dataTraining}[0]{\mathcal{D}_{\text{train}}}
\newcommand{\numTrain}[0]{N_{\text{train}}}

\newcommand{\fbtu}[0]{f_{\text{BTU}}}
\newcommand{\fpred}[0]{f_{\text{pred}}}
\newcommand{\Uniform}[2]{\text{U}\left(#1,#2\right)}

\newcommand{\Lognormal}[2]{\text{lognormal}\left(#1,#2\right)}

\newcommand{\Beta}[2]{\text{Beta}\left(#1,#2\right)}

\newcommand{\Quantile}[2]{Q_{#1}(#2)}
\newcommand{\quantPrev}[1]{\Quantile{\text{prev}}{#1}}
\newcommand{\quantCuminc}[1]{\Quantile{\text{cuminc}}{#1}}
\newcommand{\quantReff}[1]{\Quantile{\text{Reff}}{#1}}
\newcommand{\logten}[1]{\log_{10}(#1)}
\newcommand{\bdsky}[0]{BDSky}

\newcommand{\depthBTU}[0]{2}
\newcommand{\widthBTU}[0]{50}
\newcommand{\activationBTU}[0]{ELU}
\newcommand{\depthPred}[0]{2}
\newcommand{\widthPred}[0]{100}
\newcommand{\activationPred}[0]{ELU}

\newcommand{\dataNumTrain}[0]{8,000}

\newcommand{\dataNumTest}[0]{1,000}
\newcommand{\batchSize}[0]{32}
\newcommand{\batchWidth}[0]{128}
\newcommand{\numEpochBig}[0]{500}
\newcommand{\numEpochSmall}[0]{250}
\newcommand{\learningRate}[0]{10^{-3}}
\newcommand{\dropoutPerc}[0]{10}

\newcommand{\compTimeNBE}[0]{0.167}
\newcommand{\compTimeMCMC}[0]{23.3}

\newcommand{\Eqref}[1]{Eq.~\eqref{#1}}
\newcommand{\Secref}[1]{\S\ref{#1}}
\newcommand{\SISecref}[1]{\S SI\ref{#1}}
\newcommand{\Figref}[1]{Fig.~\ref{#1}}
\newcommand{\SIFigref}[1]{Fig.~SI\ref{#1}}
\newcommand{\Tabref}[1]{Tab.~\ref{#1}}
\newcommand{\SITabref}[1]{Tab.~SI\ref{#1}}

\usepackage{xcolor}
\usepackage[shadow,loadshadowlibrary]{todonotes}

\definecolor{lightGreen}{HTML}{edf8e9}
\definecolor{darkGreen}{HTML}{006d2c}

\definecolor{lightBlue}{HTML}{edefff}
\definecolor{darkBlue}{HTML}{0019fc}

\begin{document}

\maketitle

\begin{abstract}
  Phylodynamics is used to estimate epidemic dynamics from phylogenetic
  trees or genomic sequences of pathogens, but the
  likelihood calculations needed can be challenging for complex
  models. We present a neural Bayes estimator (NBE) for key epidemic
  quantities: the reproduction number, prevalence, and cumulative
  infections through time. By performing quantile regression over tree
  space, the NBE allows us to estimate posterior medians and credible intervals
  directly from a reconstructed tree. Our approach uses a recursive
  neural network as a tree embedding network with a prediction network
  conditioned on time and quantile level to generate the estimates. In
  simulation studies, the NBE achieves good predictive performance,
  with conservative uncertainty estimates. Compared with a BEAST2
  fixed-tree analysis, the NBE gives less biased estimates of
  time-varying reproduction numbers in our test setting. Under a
  misspecified sampling model, the NBE performance degrades (as
  expected) but remains reasonable, and fine-tuning a pre-trained
  model yields estimates comparable to those from a model trained from
  scratch, at substantially lower computational cost.
\end{abstract}

\section{Introduction}


In this paper we consider the birth-death-sampling model of an epidemic: infections generate (give birth to) new
lineages, infected individuals become uninfectious and a lineage
ends (death), and a subset of these ``becoming uninfectious events''
involve sequencing of the pathogen (sampling). The observed (sampled) infections then form the tips of a reconstructed viral phylogeny (hereafter the tree).

\begin{figure}[ht!]
  \centering
  \includegraphics[width=0.5\linewidth]{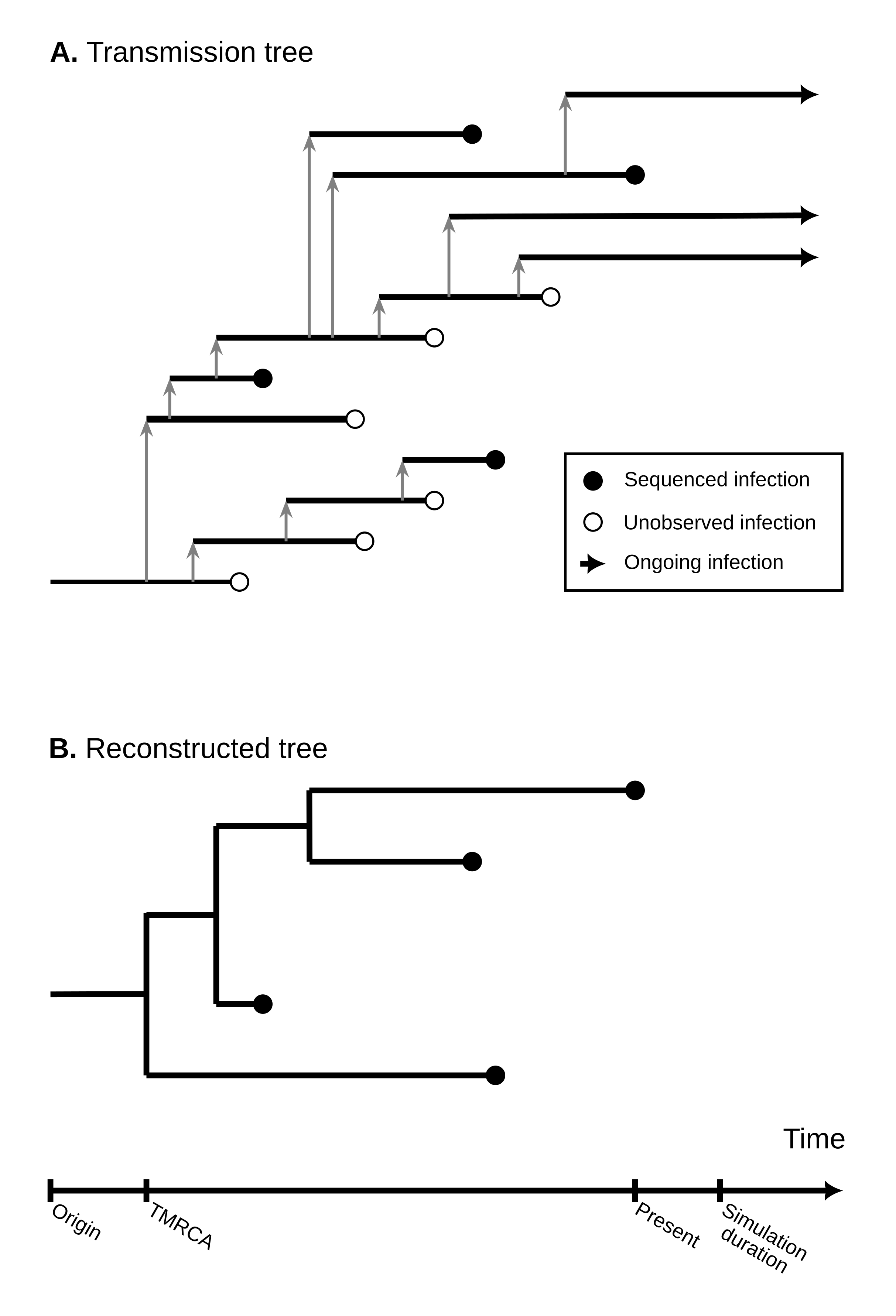}
  \caption{\label{fig:timelines} The transmission process is modeled
    as a birth-death process, in which the transmission of the
    pathogen is represented by the birth of a new lineage, and the end
    of an infectious period is represented by the death of a lineage.
    \textbf{A.} The transmission tree is a complete description of the
    transmission and observation processes. Starting from a single
    infectious individual, new infections are indicated with grey arrows,
    and an infection may end with the pathogen genome sequenced (filled dot). At
    the end of this example, there is a prevalence of three ongoing
    infections and the cumulative number of infections is 13. \textbf{B.} The
    reconstructed tree describes the connections between the sequenced
    infections. The time of the final sequenced infection is
  designated as the present.}
\end{figure}

In \Figref{fig:timelines} we illustrate the difference between the
(latent) \emph{transmission tree} and the \emph{reconstructed tree}
obtained from a set of sequenced samples. Our goal is to estimate
time-varying epidemic quantities from the reconstructed tree, in
particular the effective reproduction number, $\reff$; the common logarithm of
the prevalence of infection, $\logten{\numPrevX}$ (where \(\numPrevX\) is the
  number of extant lineages in the transmission tree at a given
time); and the common logarithm of the cumulative number of infections,
$\logten{\numCumX}$ (where \(\numCumX\) is the number
of transmission events prior to a given time). These quantities are
central for understanding transmission, evaluating control
measures, and contextualizing genomic surveillance data.

Likelihood-based Bayesian phylodynamic methods (e.g.\ those typically performed with BEAST X~\citep{baele2025beastx} or BEAST2~\citep{bouckaert2019beast})
provide a principled framework for joint inference of trees and
epidemiological parameters, but often require evaluating complex likelihoods
which hampers progress \citep{voznica2022deep}. In practice, inference
is often limited by the cost of repeatedly evaluating complex
likelihoods and exploring high-dimensional posteriors via MCMC.
Moreover, scaling to large genomic datasets may require careful
approximation, strong model simplifications or substantial
computational resources~\citep{zarebski2022computationally,zarebski2025estimating}.

Recent work has explored \emph{amortized} methods based on deep
learning, as an alternative to MCMC based
approaches~\citep{zammitmangion2025neural}. In neural approaches to amortized
inference a neural network is trained to map data to estimates.
Since the resulting models are usually
fast to evaluate they can rapidly generate estimates for new datasets.
For example, \cite{voznica2022deep} introduces \emph{PhyloDeep} for
parameter estimation using a fixed-size vectorization: the compact
bijective ladderized vector (CBLV). To avoid identifiability issues,
\cite{voznica2022deep} provide the sampling probability as input to
the network. An alternative to the CBLV is
\emph{Phylo2Vec}~\citep{penn2024phylo2vec} tree embedding which could be extended to account for branch lengths.

The approach of representing trees using hand-crafted encodings has
some limitations. Often lineages that are close in the tree may be
encoded in distant parts of the encoding, potentially diluting local
structure across a large receptive field; it imposes a bounded input
size, requiring padding for smaller trees and splitting or subsampling
for larger trees; and because estimation operates on a fixed
vectorization, extending the method to other tasks, such as labeling
internal nodes, is non-trivial. Moreover, much of the existing
deep-learning work in this area has focused on models with constant
rates, and uncertainty quantification is frequently obtained via
variants of the parametric bootstrap, which requires additional
simulation at prediction time \citep{voznica2022deep,lambert2023deep}.


The literature mentioned above suggests several practical gaps for amortized,
simulation-based estimators in phylodynamics. First, many approaches
rely on fixed-size tree encodings, which complicate the use of
variable-size trees and the inclusion of additional metadata. Second,
quantities that change over time (such as \(\refft{t}\)) are
often of interest, for example, to assess the impact of particular
interventions. However, estimation of time-varying quantities have
received less attention in deep-learning-based approaches. Third,
efficient uncertainty quantification remains challenging: many methods
require additional simulation (e.g.\ bootstrap-style procedures) to
obtain intervals, which reduces the computational advantages of
amortization. Finally, applied work frequently needs multiple outputs
(e.g.\ reproduction number, prevalence, and cumulative incidence),
ideally inferred in a single, consistent framework.

Other architectures have been proposed, including graph neural networks (GNNs)
and recurrent neural networks. Comparisons have highlighted that GNNs can perform poorly in phylodynamic settings
if not configured appropriately \citep{lajaaiti2023comparison}, motivating tailored pooling schemes
and architectural refinements \citep{leroy2025graph}. Taken together,
these developments motivate a need for methods that (i) preserve tree
structure, (ii) naturally extend to trees of any size, (iii) support
time-varying quantities, and (iv) provide calibrated uncertainty with minimal
additional computation.


We address these gaps with a neural Bayes estimator (NBE) trained
via simulation-based inference under a birth-death-sampling generative
model. The estimator takes as input a reconstructed phylogeny $\tree$
and an estimate of the mean infectious duration $\sigma^{-1}$, and
returns \emph{quantiles} of the posterior distribution of $\refft{t}$,
$\logten{\numPrev{t}}$, and $\logten{\numCum{t}}$ at user-specified times $t$ and
quantile levels $\tau\in[0,1]$. A single trained model can
produce posterior marginal medians and credible intervals without
running MCMC or performing additional bootstrap simulations at
prediction time.

The approach we take to learning a useful embedding of trees
makes use of a recursive neural network (RvNN),
\citep{frasconi1998general}. The RvNN generates tree embeddings by
propagating information from tips to the root, producing a
fixed-length embedding of the full tree. This embedding is then
combined with a feed-forward prediction network along with the time
and quantile level to produce the desired estimates. As we demonstrate
below, separating the networks into an embedding and a prediction
network enables the embedding component to be re-used in other
estimation problems with only minor fine-tuning. We assume the
reconstructed tree is observed without error and that $\sigma$ is
available (or can be estimated externally), which is not unusual in
applied work. For example a tree may be obtained via maximum
likelihood or parsimony methods and clinical information about
infectious periods, such as the mean generation time, may be available.


Our RvNN architecture annotates each internal node with an
embedding of its descendant lineages, avoiding ad hoc padding or
splitting and offering substantial flexibility. Beyond learning a
useful embedding, we incorporate quantile regression directly into the
amortized estimator, enabling prediction of arbitrary posterior
quantiles (and hence credible intervals) without additional
simulation. Finally, by conditioning on time $t$ we obtain a single
model that can be queried to produce trajectories of $\refft{t}$,
$\logten{\numPrev{t}}$, and $\logten{\numCum{t}}$ across the epidemic.


In \Secref{sec:methods} we describe the NBE and how it was trained: in
\Secref{subsec:bds-model} we describe the birth-death-sampling model
used to simulate the data, in \Secref{subsec:nn-arch} we describe the
neural network architecture used in the NBE to generate estimates, and
in \Secref{sec:training-and-simulation} we describe how the NBE was
trained. In \Secref{sec:results} we report the results of
computational experiments: in \Secref{subsec:results:calibration} we
describe a simulation study assessing the calibration of the NBE's
estimates, in \Secref{sec:mcmc-comparison} we compare the NBE's
estimates of the reproduction number to those generated with using
MCMC, specifically a BEAST2 analysis using \bdsky, and in
\Secref{sec:results-sensitivity} we carry out a sensitivity analysis
of how the NBE performs when the sampling model is misspecified, and
demonstrate how fine-tuning offers a solution to overcoming issues of
this nature. \Secref{sec:discussion} offers some discussion on the use
of NBEs in phylodynamic inference.

\section{Methods}\label{sec:methods}

\subsection{Birth-death-sampling model}\label{subsec:bds-model}

The generative model we consider is a birth-death-sampling model
\citep{stadler2024decoding} with (random) piece-wise constant
functions used for the parameters for the reproduction number,
\(\reff\), and the proportion of infections that end in sequencing,
\(p_{\psi}\); and a (randomly sampled) constant for the net rate of
becoming uninfectious, \(\sigma\). Further details of the generative
model are provided in \SISecref{sec:si:simulation}. The prior
distribution over the parameters is detailed in \SITabref{tab:si:prior-reduced}; the
prior is based on one used for an analysis of SARS-CoV-2 transmission
\citep{douglas2021phylodynamics}.

We assume that a tree, \(\tree\), can be reconstructed without error from the
sequenced samples in the simulation, and that the average duration of being
infectious, or equivalently, the net becoming-uninfectious rate, \(\sigma\) is known.
Given an observation
of \((\tree,\sigma)\), we are interested in the posterior distribution
of three quantities throughout the course of the epidemic: the
reproduction number, \(\reff\), the common logarithm of the prevalence of infection,
\(\logten{\numPrevX}\), and the common logarithm of the cumulative number of infections, \(\logten{\numCumX}\).
Since the prevalence and cumulative number of infections can vary over
several orders of magnitude, we consider the common logarithm (with
base \(10\)) of these quantities rather than their value directly.

\subsection{Neural Bayes estimator}\label{subsec:nn-arch}

Neural amortized inference methods have been developed to avoid
running expensive likelihood-based MCMC analyses from scratch for every new
dataset~\citep{zammitmangion2025neural}. Neural Bayes estimators (NBEs) use
neural networks to approximate Bayes
estimators~\citep{sainsburydale2023likelihood}. NBEs, and similar methods, use
simulated data to train a neural network that
maps an observed tree directly to parameter estimates~\citep{voznica2022deep}.
Our neural Bayes estimator (NBE) performs quantile regression over
tree space for the marginal posterior quantities of interest. The
required function is
\(f:\treeSpace\times\mathbb{R}_{>0}\times\mathbb{R}_{\geq0}\times[0,1]\longrightarrow\mathbb{R}^{3}\)
such that
\begin{equation*}
  f(\tree,\sigma^{-1},t,\tau)=\left(\quantPrev{\tau},\quantCuminc{\tau},\quantReff{\tau}\right)
\end{equation*}
where \(\tree\in\treeSpace\) is the tree reconstructed from the
sequenced samples, \(\sigma^{-1}\in\mathbb{R}_{\geq0}\) is the average
duration of infectiousness, \(t\in\mathbb{R}_{\geq0}\) indicates the
time in the process we are interested in, and \(\tau\in[0,1]\)
indicates the desired quantile. The \(\quantPrev{\tau}\),
\(\quantCuminc{\tau}\), and \(\quantReff{\tau}\) are the quantile
functions of the posterior (predictive) distributions of the
(common logarithm of the) prevalence and cumulative incidence, and reproduction number. Given \(f\)
we can compute posterior medians and marginal credible intervals for
\(\logten{\numPrev{t}}\), \(\logten{\numCum{t}}\) and \(\refft{t}\).


Since we cannot obtain the function \(f\) directly, we will
approximate it with a neural network \(f_{\text{NBE}}\):
\begin{equation*}
  f(\tree,\sigma^{-1},t,\tau)\approx f_{\text{NBE}}(\tree,\sigma^{-1},t,\tau) = \fpred(\fbtu(\tree),\textsc{height}(\tree),\sigma^{-1},t,\tau)
\end{equation*}
\noindent
where \(\fbtu\) and \(\fpred\) are two neural networks as described
below (in \Secref{sec:btu} and \Secref{sec:pred-unit} respectively),
and \(\textsc{height}\) returns the tree's height. The parameters of
\(f_{\text{NBE}}\) are the parameters of the subnetworks \(\fbtu\) and
\(\fpred\) described below.


\subsubsection{Binary tree unit}\label{sec:btu}

The function \(\fbtu\) is the \emph{binary tree unit} (BTU), a
recursive neural network (RvNN) \citep{frasconi1998general}.
RvNN describe functions on directed acyclic graphs of bounded degree and size, but we only consider the special case of
binary trees with weighted edges. Here we consider the definition of a \(\fbtu\) which
embeds a tree \(\tree\) into \(\mathbb{R}^{n}\) for a given \(n\).

Before defining \(\fbtu\), it is helpful to define some functions on
the set of trees. Note that for some of these functions, (e.g.
\(\textsc{depth}\)), there is an implicit supertree that we have
omitted to simplify the presentation. Let \(\treeSpace\) denote the
set of binary trees (with branch lengths) and
\(\treeSpace_{1}\subset\treeSpace\) the subset of trees which are a
single leaf.
\begin{itemize}
  \item \(\textsc{left}~(\text{and}
    ~\textsc{right}):\treeSpace/\treeSpace_{1}\to\treeSpace\) extracts
    the left (or right) subtree of a given tree
  \item \(\textsc{branch}:\treeSpace\to\mathbb{R}_{\geq0}\) extracts the
    length of the branch connecting a tree to its parent, for a root
    this is defined to be zero.
  \item \(\textsc{depth}:\treeSpace\to\mathbb{R}_{\geq0}\) returns the
    depth of the root node of the given tree (within a supertree).
  \item \(\textsc{height}:\treeSpace\to\mathbb{R}_{\geq0}\) returns the
    height of a given tree.
\end{itemize}
\noindent
With these functions we can then define
\(\fbtu:\treeSpace\to\mathbb{R}^{n}\) in the following way:
\begin{equation}\label{eq:btu}
  \fbtu(\tree) =
  \begin{cases}
    \begin{aligned}
    &\sigma([\textsc{depth}(\tree) / h_{\text{super}}, \textsc{branch}(\tree) / h_{\text{super}}, \mathbf{0}]), \quad\quad\text{if } \tree\in\treeSpace_{1} \\[3mm]
      &g([\sigma([\textsc{depth}(\tree) / h_{\text{super}}, \textsc{branch}(\tree) / h_{\text{super}}]), \quad\quad\text{otherwise} \\
        & \quad \fbtu(\textsc{left}(\tree)), \\
        & \quad \fbtu(\textsc{right}(\tree))])
    \end{aligned}
  \end{cases}
\end{equation}
\noindent
where \(g:\mathbb{R}^{2n+2}\to\mathbb{R}^n\) is a neural network with
activation function \(\sigma\), \(\mathbf{0}\) is a vector of \(n-2\)
zeros and \(h_{\text{super}}\) is the height of the supertree for the
given tree. We used a multilayer perceptron (MLP) for \(g\) with
\depthBTU~hidden layers of width \widthBTU~(i.e.\ we are embedding trees into \(\mathbb{R}^{50}\)) with the
exponential linear unit (\activationBTU)~activation function
\citep{clevert2015fast}.

In the definition of \(\fbtu\), we normalized both the node depth
and branch lengths by the supertree height to ensure they take values from
zero to one. Due to the scaling of branch lengths and depths by the
height of the tree the BTU learns an embedding of trees that is
invariant of the scaling of the tree. This is important as it means
that the embedding does not depend upon the \emph{absolute} time
scale.


\subsubsection{Prediction unit}\label{sec:pred-unit}

The function \(\fpred\) is the prediction unit,
\(\fpred:\mathbb{R}^{n}\times\mathbb{R}_{>0}\times\mathbb{R}_{>0}\times\mathbb{R}_{\geq0}\times[0,1]\to\mathbb{R}^{3}\),
defined as
\begin{equation}\label{eq:pred}
  \fpred(r_{\tree},t_{h},\tilde{\sigma}^{-1},\tilde{t}_{m},\tau) = h([r_{\tree},t_{h},\tilde{\sigma}^{-1},\tilde{t}_{m},\tau])
\end{equation}
\noindent
where \(h:\mathbb{R}^{n+4}\to\mathbb{R}^3\) is a neural
network returning the quantiles for \(\reff\), \(\logten{\numPrevX}\), and \(\logten{\numCumX}\); \(r_{\tree}=\fbtu(\tree)\) is the embedding of the
reconstructed tree from the BTU;
\(t_{h}=\textsc{height}(\tree)\) is the height of the reconstructed
tree (because there needs to be some time scale information);
\(\tilde{\sigma}^{-1}\) is the average infection duration scaled by
the height of \(\tree\); and \(\tilde{t}_{m}\) is the measurement
time, again scaled by the height of the input tree.
We used an MLP for \(h\) with \depthPred~hidden layers of width \widthPred~and
\activationPred~activations. When generating the predictions, we applied a
softplus activation to get the \(\reff\) estimate, to ensure it is positive.

\subsection{Data and training}\label{sec:training-and-simulation}

\subsubsection{Loss function}\label{sec:loss-fn}

To learn the quantiles of the posterior distribution we used the
\emph{pinball loss function} (a.k.a. the \emph{tilted absolute value
loss function}) \citep{zammitmangion2025neural}, for a given quantile
\(\tau\), the pinball loss \(\ell_{\tau}\) is given by
\begin{equation*}
  \ell_{\tau}(\hat{y}, y) =\begin{cases}
  \tau (y - \hat{y}) & \text{if } y \geq \hat{y} \\
  (1 - \tau) (\hat{y} - y) & \text{otherwise}.
  \end{cases}
\end{equation*}
To learn all of the quantiles simultaneously, we take the expected
loss summed over the three of the quantities of interest.
We approximate the expectation with the Monte Carlo estimate:
\begin{equation}\label{eq:loss-fn}
\mathbb{E}_{y,\tau}[\ell_{\tau}(\hat{y},y)]\approx\frac{1}{IJ}\sum_{I=1}^{I}\sum_{j=1}^{J}\sum_{k=1}^{3}\ell_{\tau_{i}}(\hat{y}_{i,j,k},y_{i,j,k})
\end{equation}
\noindent
for a batch of \(I\) simulations, each of which has \(J\)
measurements, where \(\tau_{i}\) and \(y_{i,j,k}\) are sampled from
the prior distribution with \(y_{i,j,k}\) the measurement of the
\(k\)th quantity at the \(j\)th observation time in the \(i\)th
simulation.


\subsubsection{Optimization}\label{sec:training-alg}

The NBE was optimized using AdamW \citep{loshchilov2017decoupled}
using the default parameters in PyTorch (e.g. a learning rate of
\(\learningRate\)). We used batches of \batchSize~simulations, each with
measurements at \batchWidth~points in time throughout the
simulation. For each batch, we used a different value of
\(\tau_{i}\sim\Beta{0.5}{0.5}\) in \Eqref{eq:loss-fn}. We used simple
early stopping, selecting the state of the model with the lowest
validation error across the training epochs. All model weights
and biases were initialized using the default method in PyTorch,
excluding the last layer of the prediction unit where we initialized
the bias to match the sample average of the training data and set the
weights to zero. During training, we used dropout on internal
layers with \(\dropoutPerc\%\) of nodes zeroed out. All of the network
optimization was carried out on CPUs comparable to what would be available in a mid-range desktop.

\subsubsection{Simulating training and testing data}\label{sec:simulation}

The training data for our NBE was simulated from the model described
above. A more detailed description of the data generation process is given
in \SISecref{sec:si:simulation}. Each simulation produces a
reconstructed tree of sampled infections, a value of the net
becoming-uninfectious rate, and measurements (through time) of the
three quantities. I.e.\ at time \(t\) in the simulation, we measure
\((\refft{t},\logten{\numPrev{t}},\logten{\numCum{t}})\). These measurements are taken
at \batchWidth~randomly sampled times between the beginning of the process and the randomly sampled stopping time. The
training dataset has \dataNumTrain~simulations, with a further
\dataNumTest~each for validation and testing.




\FloatBarrier

\section{Results}\label{sec:results}

\subsection{Simulation study}\label{subsec:results:calibration}

\begin{figure}[ht!]
  \centering
  \includegraphics{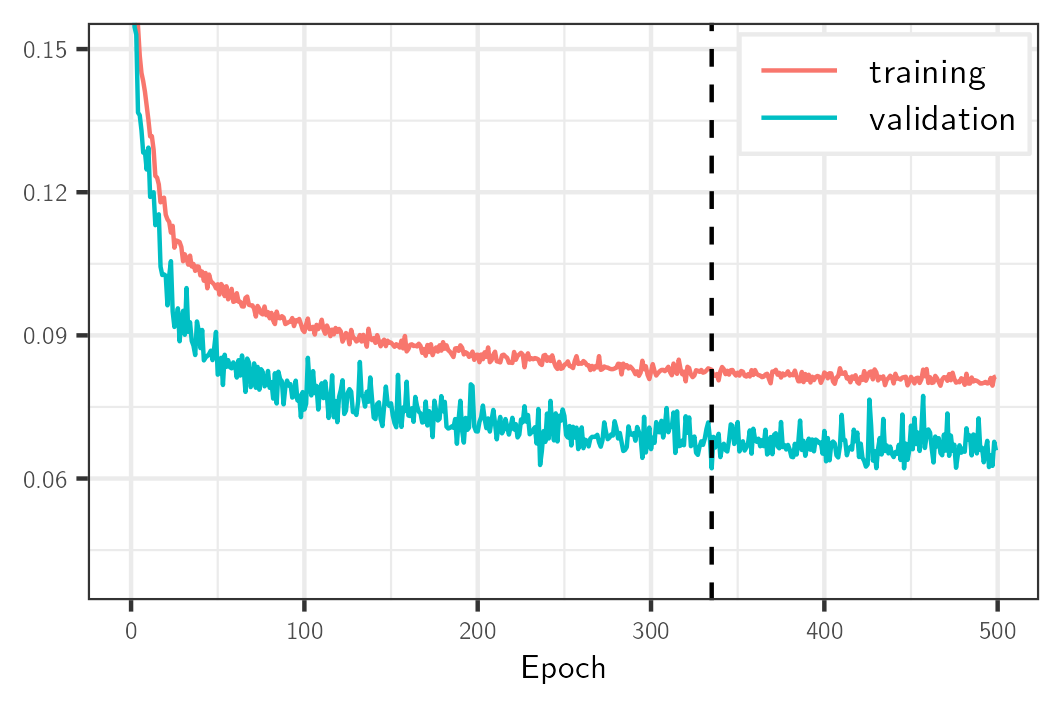}
  \caption{\label{fig:si:learning-curves} Learning curves show that the
    training and validation losses across epochs have
    effectively converged by epoch 500. The validation loss is less
    than the training loss because drop-out was applied when assessing
    the training loss. The vertical dashed line indicates the epoch
  with the lowest validation loss.}
\end{figure}

Using the simulation and training procedure described above, we
trained an NBE to predict time-varying epidemic quantities from
reconstructed trees. In \SIFigref{fig:si:learning-curves} we show the training and validation loss
throughout training. We use the model weights with the lowest validation
loss across the \numEpochBig~training epochs. Below we describe our assessment of the predictive performance
and calibration of this trained model on the held-out testing simulations.

\begin{figure}[ht!]
  \centering
  \includegraphics[width=0.99\linewidth]{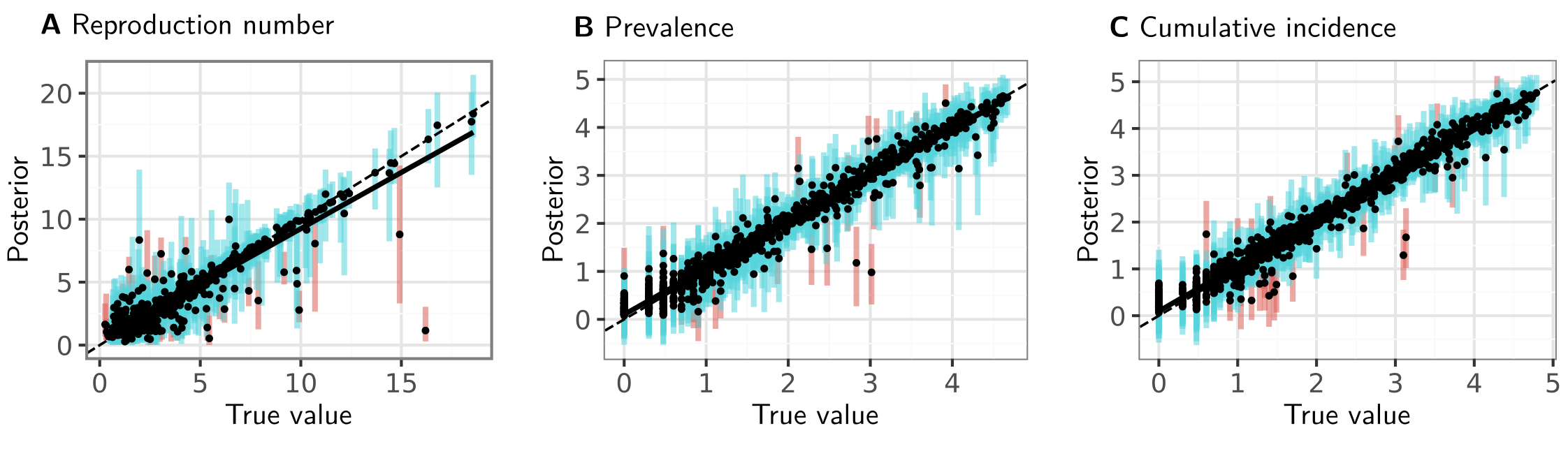}
  \caption{\label{fig:testing-performance}The predictions and \(95\%\)
    confidence intervals on the testing data for the three prediction
    targets \textbf{A} the reproduction number, \textbf{B} the (common logarithm of the)
    prevalence of infection, and \textbf{C} the (common logarithm of the) cumulative number of
    infections. The bars associated with each point are blue if the
    estimated credible interval contains the true value and red
    otherwise. The dashed black line indicates perfect agreement and
  the solid black line is a least squares fit.}
\end{figure}

In \Figref{fig:testing-performance} we show the NBE estimates (and
95\% credible intervals) of the reproduction number, \(\reff\), and
the common logarithm of prevalence, \(\logten{\numPrevX}\), and
cumulative incidence, \(\logten{\numCumX}\), for a
randomly sampled time in each of the \dataNumTest~testing
simulations. The estimates of the prevalence of infection and the
cumulative number of cases is very good (over 95\% of the variance is
explained with little bias). On average the estimates for the
reproduction number are good (approximately 90\% of the variance is
explained with little bias).

In \Tabref{tab:derp-summary} we summarize the performance across the
\batchWidth~measurements for each of the \dataNumTest~testing
trees. The
uncertainty in the predictions of the reproduction number appears to
be well-calibrated, if slightly pessimistic: the credible intervals, in
particular the 50\% credible intervals,
are too large. As an indication of the range of values expected of a
well-calibrated estimate, in a hypothesis test with a $0.05$ level of
significance, for the 50\% intervals, the acceptable range would cover
0.469--0.531; and for the 95\% intervals, the acceptable range would
cover 0.936--0.963.

\begin{table}[ht]
  \centering
  \begin{tabular}{lccccc}
    \hline
    \textbf{Quantity} & \multicolumn{1}{c}{\textbf{$R^2$}} & \multicolumn{1}{c}{\textbf{Bias}} & \multicolumn{1}{c}{\textbf{Cover. $50\%$}} & \multicolumn{1}{c}{\textbf{Cover. $95\%$}} \\ \hline
    Reproduction number,  \(\reff\)              & 0.881  & -0.020  & 0.690 & 0.959   \\
    Prevalence,           \(\logten{\numPrevX}\) & 0.966  & 0.047  & 0.606 & 0.968   \\
    Cumulative incidence, \(\logten{\numCumX}\)  & 0.971  & 0.045  & 0.604 & 0.966   \\ \hline
  \end{tabular}
  \caption{\label{tab:derp-summary} Summary statistics for the estimates from the NBE over \dataNumTest~trees, each quantity being measured at \batchWidth~different times throughout the simulation. \(R^{2}\) is the coefficient of determination, the proportion of the variance explained. Coverage at X\% indicates the proportion of X\% credible intervals that contained the true value from the simulation.}
\end{table}

\subsection{Comparison to existing method}\label{sec:mcmc-comparison}

\begin{figure}[ht!]
  \centering
  \includegraphics[width=1.0\linewidth]{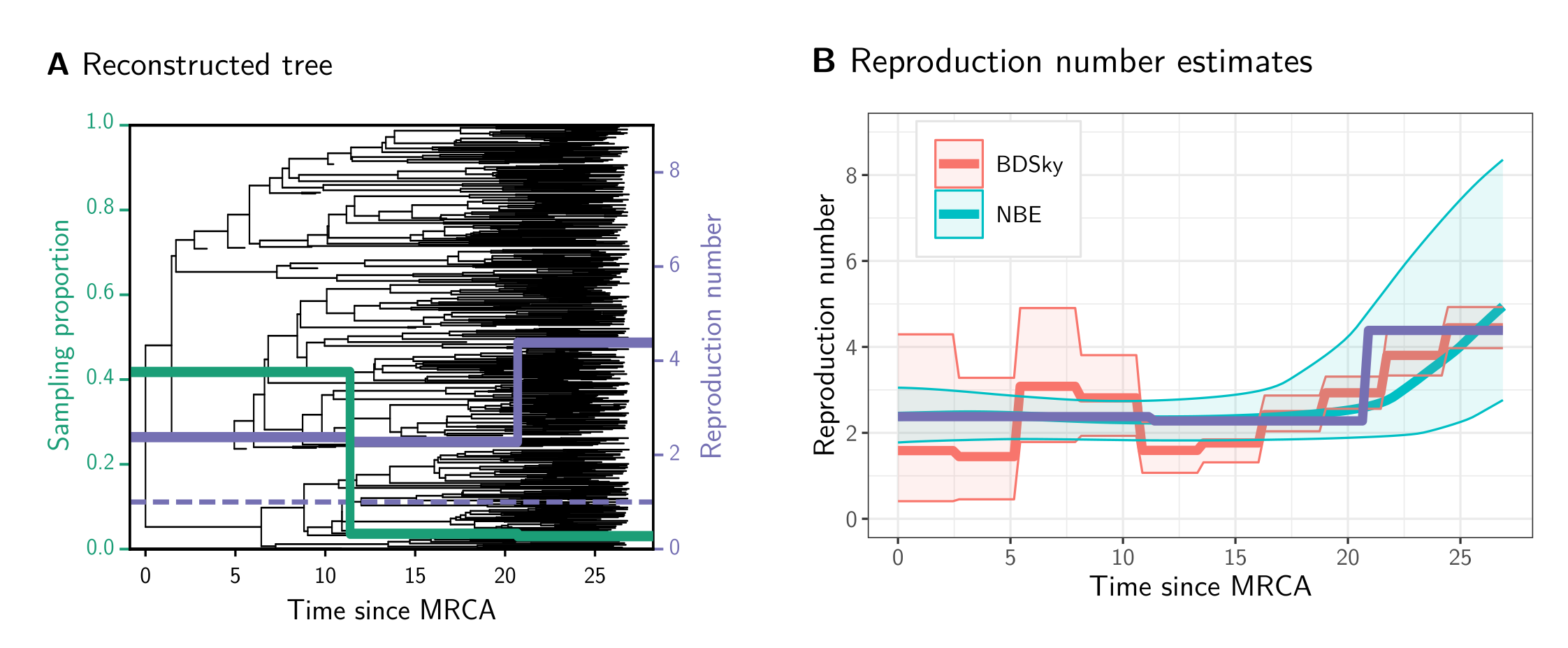}
  \caption{\label{fig:example} Example of a realization of the
    birth-death sampling process and the estimates of the reproduction
    number generated for this data using both MCMC and NBE.
    \textbf{A.} The simulated tree and the varying reproduction number
    and proportion of infections sampled through time. \textbf{B.} Estimates of the effective reproduction
    number produced by BDSky MCMC and the NBE and their \(95\%\) credible
    intervals along with the true value. The NBE estimates do not make
    a piece-wise constant assumption, so they vary smoothly through
  time, whereas the MCMC uses a piece-wise constant estimator. }
\end{figure}

Using the same trained NBE considered in
\Secref{subsec:results:calibration}, we carried out a comparison of
the NBE's estimates with those generated by MCMC with BEAST2 using the
BDSky tree prior with a fixed (known) tree. In \Figref{fig:example} we
show an example of the estimate of the effective reproduction number
through time generated with the NBE and the corresponding estimates
generated with the MCMC methods.
The BDSky tree prior used in the MCMC is misspecified. The simulated
testing data uses a randomly sampled piece-wise constant function with
two or three pieces that change at random times for both the
reproduction number and the sampling proportion; the model in the MCMC
assumes a constant sampling proportion and a piece-wise constant
reproduction number that changes at 9 points.

To compare the performance of our neural estimator to that of a
gold-standard BDSky MCMC method, we considered the performance of a
fixed-tree \bdsky~analysis~\citep{stadler2013birth} with BEAST2~\citep{bouckaert2019beast} across the
\dataNumTest~testing simulations. Details of the MCMC analysis are provided
in the \SISecref{sec:si:bdsky}.

In \Tabref{tab:beast-comparison} we show the results of this comparison:
the NBE explains more of the variability (as measured by the coefficient of
determination, \(R^{2}\)) and is less biased.
The coverage of the credible intervals is greater for the NBE than the BDSky MCMC. At
both 50\% and 95\% the NBE coverage exceeds the desired coverage, while for the
MCMC it is smaller than the desired level. This is likely due to the model
misspecification between the model used in the MCMC and the model used to
simulate the testing data. Note
that a small proportion of estimates are missing from the MCMC analysis because
if a measurement time falls before the TMRCA or after the most recent
sample (see \Figref{fig:timelines}) then there is no estimate
generated for that measurement.

\begin{table}[ht]
  \centering
  \begin{tabular}{lccccc}
    \hline
    \textbf{Method} & \multicolumn{1}{c}{\textbf{$R^2$}} & \multicolumn{1}{c}{\textbf{Bias}} & \multicolumn{1}{c}{\textbf{Cover. $50\%$}} & \multicolumn{1}{c}{\textbf{Cover. $95\%$}} & \multicolumn{1}{c}{\textbf{Time (sec)}} \\ \hline
    NBE    & \textbf{0.881} & \textbf{-0.020} & 0.690          & \textbf{0.959} & \textbf{\compTimeNBE}     \\
    MCMC   & 0.763          & -0.348          & \textbf{0.464} & 0.906          & \compTimeMCMC             \\ \hline
  \end{tabular}
  \caption{\label{tab:beast-comparison} Comparison of estimates of the
    reproduction number through time generated by the NBE and a BEAST2
    MCMC analysis using BDSky. These metrics are computed from the analysis of the
    estimates at \batchWidth~times in the \dataNumTest~testing simulations using both the NBE and MCMC (with BEAST2). The values in Time for the MCMC analysis are the average number of seconds required to attain an effective sample size (ESS) of 200 for all variables. The NBE
    outperforms the MCMC on most of the considered metrics,
  (\textbf{bold} indicates the best value for each metric.)}
\end{table}



\subsection{Sensitivity analysis}\label{sec:results-sensitivity}

The practical utility of an amortized estimator depends on
its ability to generalize beyond the precise assumptions used to
generate the training data. We evaluate sensitivity to model
assumptions in two ways: in \Secref{results:model-misspecification}, we consider
misspecification of the observation model; and in
\SISecref{sec:si:prior-sensitivity}, we report qualitatively similar results from an additional analysis
of sensitivity to prior misspecification.

\subsubsection{Model misspecification}\label{results:model-misspecification}

To understand the NBE's performance under model misspecification, we
simulated a second, alternative dataset. This alternative dataset
(described in more detail in \SISecref{sec:si:alt-model}) is simulated from a process in which initially there is no sampling, and then after a random time sampling is activated.
This is a plausible scenario to consider, for example, it could correspond to
an initial period in which the pathogen is
spreading before it has come to the attention of the surveillance
system. As in \Secref{sec:simulation}, we split this alternative data
into training, validation and testing datasets.

Using both this new dataset, and the original one described above, we trained
the three models described below, in each case running the optimizer as
described above for \numEpochSmall~epochs: \textbf{zero-shot}, in which we
take the model trained on the original training dataset for 250 epochs without modification;
\textbf{fine-tuning}, in which we take the model trained on the original training
dataset (i.e. the zero-shot model) and fine-tune the prediction unit on the alternative training dataset; and
\textbf{random initialization}, in which we train a model from random
initialization on the alternative training dataset (using the alternative
validation dataset for early stopping).

In \Tabref{tab:sensitivity} we summarize the performance of the three
models on the alternative testing dataset. The first row demonstrates
that the fine-tuning process is several orders of magnitude faster
than training the full model. The performance of the fine-tuned model
is very similar to that of the model trained from random initialization.

\begin{table}[ht]
  \centering
\begin{tabular}{lrrrr}
\hline
     & \multicolumn{1}{l}{} & \multicolumn{1}{l}{\begin{tabular}[c]{@{}l@{}}Zero-shot\\ pre-trained\end{tabular}} & \multicolumn{1}{l}{\begin{tabular}[c]{@{}l@{}}Fine-tuned\\ pre-trained\end{tabular}} & \multicolumn{1}{l}{\begin{tabular}[c]{@{}l@{}}Randomly\\ initialized\end{tabular}} \\ \hline
Training time          & seconds              & \textbf{0}            & \underline{215}             & 242978            \\ \hline
\multirow{2}{*}{\begin{tabular}[c]{@{}l@{}}Reproduction\\ number\end{tabular}}     & $R^2$ & 0.692           & \textbf{0.754}            & \underline{0.742}           \\
        & 95\% cover.      & 92           & \underline{94}            & \textbf{95}           \\ \hline
\multirow{2}{*}{\begin{tabular}[c]{@{}l@{}}Prevalence of\\ infection\end{tabular}} & $R^2$ & 0.956         & \textbf{0.969}          & \underline{0.965}         \\
        & 95\% cover.      & 94         & \textbf{96}          & \textbf{96}         \\ \hline
\multirow{2}{*}{\begin{tabular}[c]{@{}l@{}}Cumulative\\ infections\end{tabular}}   & $R^2$ & 0.951           & \textbf{0.965}            & \underline{0.962}           \\
        & 95\% cover.      & 91           & \underline{96}            & \textbf{95}           \\ \hline
\end{tabular}\caption{\label{tab:sensitivity} Fine-tuning a pre-trained model is much faster than training a new model and recovers much of the performance lost to model misspecification. The coefficient of determination, $R^2$, and coverage of the 95\% credible interval is shown for each model and quantity. The \textbf{bold} entry is the best in each row, with the second best \underline{underlined}. The training time of the zero-shot model is zero because we assume there is a pre-trained model to be used in this case and fine-tuned in the fine-tuning example.}
\end{table}

\FloatBarrier

\section{Discussion}\label{sec:discussion}


Neural Bayes estimators (NBEs) can estimate important quantities of an
epidemic using a reconstructed phylogeny as data. The quantities we
consider here are the \emph{time varying} effective reproduction
number, prevalence of infection, and cumulative number of infections.
This represents an important extension of capability because standard
MCMC methods typically only estimate either the effective reproduction
number or (a quantity proportional to) the prevalence of infection,
and other simulation-based inference (SBI) methods are computationally
expensive when they are not completely intractable. Importantly, the
NBE method we present here can be trained without expensive hardware
(e.g., GPUs), and our simulation studies
(\Secref{subsec:results:calibration}) suggest the resulting estimators
have little bias and conservative uncertainty, with interval coverage
generally at or above nominal levels.

To understand how the NBEs perform relative to a gold-standard MCMC
method (e.g., a fixed-tree \bdsky~analysis with BEAST2), we performed
a simulation study comparing NBE estimates to those from MCMC
(\Secref{sec:mcmc-comparison}) for the effective reproduction number.
We found that NBEs outperform the MCMC on point-estimate accuracy and
bias in this simulation setting. Importantly, once trained, the NBEs
are at least an order of magnitude faster than the MCMC methods. This
speedup is typical with neural inference methods
\citep{zammitmangion2025neural}. We did not compare performance on the
estimation of prevalence and cumulative infections because there is no
definitive standard method for this problem yet.

The ability of neural networks to generalize beyond their training
data is a major concern in their practical application. We performed a
sensitivity analysis (\Secref{sec:results-sensitivity} and \SISecref{sec:si:prior-sensitivity}), to assess the
NBE's performance under model misspecification. The main observations
of this experiment are that, as expected, performance degrades with a
misspecified model, however the NBE still generates reasonable
estimates, and, with a small amount of fine-tuning (which is
computationally cheap), a pre-trained model can achieve comparable
performance to one that was trained from scratch using data simulated
from the true model. This observation supports the notion that if one
wanted to use a prior or model that differed substantially from the
one the NBE was trained with, this could be done with little
additional computational cost. This observation is important as it
builds confidence these neural estimators will perform as desired when
used on real data and are extensible to novel settings.

An additional observation related to the fine-tuning simulation studies
(\Secref{sec:results-sensitivity} and \SISecref{sec:si:prior-sensitivity}) is that the good performance of the
fine-tuned pre-trained model suggests the NBE has learned a useful
tree embedding. This is significant as it suggests the NBE could be
used as a foundation model, reducing the computational cost associated
with training subsequent NBEs for phylodynamic estimation.



NBEs offer a conceptually simple way to estimate a range of
quantities, without needing to derive complicated likelihoods,
algorithms to compute them, or informative summary statistics. This
enables the user to focus on their model and the questions they are
interested in answering, rather than the complexity of likelihood
derivations.

While NBEs have a significant cost associated with simulation and
training, the computational cost of inference is negligible. With
sensitive data increasingly stored on trusted research
environments/secure data enclaves, this allocation of computational
costs is increasingly beneficial; if necessary, the simulation and
training steps can be carried out on HPC servers and the resulting
model transferred to a lower-powered machine for inference on the
sensitive real data. In comparison, particle-based methods and ABC
place the computational cost on the inference step, and this must be
paid each time a new observed dataset is considered. As mentioned
above, the computation performed for this study was all done without
the use of an HPC cluster or specialized hardware such as GPUs,
suggesting it will be viable for many practitioners with limited
resources.


There are two main weaknesses of this work that we discuss below: the first is that the
estimators rely on a reconstructed tree as input, so the estimates do
not account for tree uncertainty; the second is that unlike
likelihood-based methods (e.g. most gold-standard MCMC approaches), it
is not obvious how these estimators will generalize beyond the scope
of their training data.

There are several ways to overcome the requirement for a reconstructed
tree. One approach is to avoid the tree entirely by using a multiple
sequence alignment (MSA) transformer \citep{rao2021msa}, which works
on an MSA directly without the need to estimate a tree --- although
this does still require an MSA to be produced, so it doesn’t entirely
solve the problem of a substantial amount of preprocessing being
involved. Overcoming the lack of uncertainty in the reconstructed tree
could be done (in a birth-death Bayesian setting) by applying the
neural estimator to a posterior sample of trees and sampling from the
implicit distributions over the prevalence and cumulative incidence to
draw a sample from the posterior predictive distribution for these
quantities. (The effective reproduction number would already have been
sampled in generating the trees in the typical analysis.)

Understanding the generalization properties of NBEs is a substantial
problem. While it is possible to force the NBE estimates to respect
certain constraints by encoding them in the neural network
architecture, this does not ensure the NBE's performance will be
maintained. Until we have a better understanding of these properties,
we advise some caution in the use of NBEs, for example, by checking
that summary statistics of the observed data fall within the
distribution of those statistics across the training data, although
there are more formal approaches to this as well, cf.,
\cite{tagasovska2019single}. Of course, model adequacy is an entirely
separate problem for which we recommend prior and posterior
simulation, which is an important part of the Bayesian workflow
\citep{gelman2020bayesian}.


There are three particularly interesting extensions to the current
work that would further enhance the utility of these methods. The
first is to reduce the main computational cost: simulating the
training data, which could be done using emulation
methods~\citep{scheurer2025uncertainty}. The second is to use
normalizing flows to estimate the full joint posterior distributions
rather than marginal predictions~\citep{radev2022bayesflow}. Finally,
the third would be to gain better diagnostics for when a particular NBE is not appropriate by detecting when a new data
point sits outside of the training data~\citep{tagasovska2019single}.

The work we present here shows recursive neural networks offer a powerful tool for representation learning of trees, and that NBEs can use reconstructed trees to quickly
estimate key epidemic trajectories with quantified uncertainty.
In our simulation studies, the NBEs are competitive with an MCMC baseline for estimating the effective
reproduction number, and NBEs remain useful under model
misspecification when fine-tuning is used. Taken together, these results support
amortized phylodynamic inference as a useful addition to traditional likelihood-based
Bayesian workflows, particularly for complex models.

\section{Data availability}

An implementation of the NBE along with the code to simulate the training data and train the neural networks are linked to from the project homepage: \url{https://aezarebski.github.io/derp/}.

\section{Acknowledgments}

AEZ acknowledges support from the Wellcome Trust [grant number 227438/Z/23/Z], and an Andrew Sisson Early Career Researcher grant.
TW acknowledges funding through the ARC Centre of Excellence for the Mathematical Analysis of Cellular Systems, and was supported by internal funding from The University of Melbourne, School of Mathematics \& Statistics.

\printbibliography{}

@article{baele2025beastx,
  author =	 "Baele, Guy and Ji, Xiang and Hassler, Gabriel W. and
                  McCrone, John T. and Shao, Yucai and Zhang, Zhenyu
                  and Holbrook, Andrew J. and Lemey, Philippe and
                  Drummond, Alexei J. and Rambaut, Andrew and Suchard,
                  Marc A.",
  title =	 {{BEAST} {X} for {B}ayesian phylogenetic,
                  phylogeographic and phylodynamic inference},
  journal =	 "Nature Methods",
  year =	 2025,
  month =	 "Aug",
  day =		 01,
  volume =	 22,
  number =	 8,
  pages =	 "1653--1656",
  doi =		 "10.1038/s41592-025-02751-x"
}

@article{bouckaert2019beast,
  author =       {Bouckaert, Remco AND Vaughan, Timothy G. AND Barido-Sottani,
                  Jo\"{e}lle AND Duch\^{e}ne, Sebasti\'{a}n AND Fourment, Mathieu AND
                  Gavryushkina, Alexandra AND Heled, Joseph AND Jones, Graham
                  AND K\"{u}hnert, Denise AND De Maio, Nicola AND Matschiner,
                  Michael AND Mendes, F\'{a}bio K. AND M\"{u}ller, Nicola F. AND
                  Ogilvie, Huw A. AND du Plessis, Louis AND Popinga, Alex AND
                  Rambaut, Andrew AND Rasmussen, David AND Siveroni, Igor AND
                  Suchard, Marc A. AND Wu, Chieh-Hsi AND Xie, Dong AND Zhang,
                  Chi AND Stadler, Tanja AND Drummond, Alexei J.},
  journal =      {PLOS Computational Biology},
  publisher =    {Public Library of Science},
  title =        {{BEAST 2.5: An advanced software platform for Bayesian
                  evolutionary analysis}},
  year =         2019,
  month =        04,
  volume =       15,
  pages =        {1-28},
  number =       4,
  doi =          {10.1371/journal.pcbi.1006650}
}

@article{clevert2015fast,
  author =	 {{Clevert}, Djork-Arn{\'e} and {Unterthiner}, Thomas
                  and {Hochreiter}, Sepp},
  title =	 {{F}ast and {A}ccurate {D}eep {N}etwork {L}earning by
                  {E}xponential {L}inear {U}nits ({ELU}s)},
  journal =	 {arXiv e-prints},
  year =	 2015,
  month =	 nov,
  doi =		 {10.48550/arXiv.1511.07289}
}

@article{zarebski2022computationally,
  doi =          {10.1371/journal.pcbi.1009805},
  author =       {Zarebski, Alexander E. AND du Plessis, Louis AND Parag,
                  Kris V. AND Pybus, Oliver G.},
  journal =      {PLOS Computational Biology},
  publisher =    {Public Library of Science},
  title =        {{A} computationally tractable birth-death model that combines
                  phylogenetic and epidemiological data},
  year =         2022,
  month =        02,
  volume =       18,
  number =       2
}

@article{zarebski2025estimating,
  author =	 {Zarebski, Alexander E. and Zwaans, Antoine and
                  Gutierrez, Bernardo and du Plessis, Louis and Pybus,
                  Oliver G.},
  title =	 {{E}stimating epidemic dynamics with genomic and time
                  series data},
  journal =	 {Journal of The Royal Society Interface},
  volume =	 22,
  number =	 227,
  year =	 2025,
  month =	 06,
  doi =		 {10.1098/rsif.2024.0632}
}

@article{loshchilov2017decoupled,
  author =	 {{Loshchilov}, Ilya and {Hutter}, Frank},
  title =	 {{D}ecoupled {W}eight {D}ecay {R}egularization},
  journal =	 {arXiv e-prints},
  year =	 2017,
  month =	 nov,
  doi =		 {10.48550/arXiv.1711.05101}
}

@article{radev2022bayesflow,
  author =	 {Radev, Stefan T. and Mertens, Ulf K. and Voss,
                  Andreas and Ardizzone, Lynton and K\"{o}the,
                  Ullrich},
  journal =	 {IEEE Transactions on Neural Networks and Learning
                  Systems},
  title =	 {{B}ayes{F}low: {L}earning {C}omplex {S}tochastic
                  {M}odels {W}ith {I}nvertible {N}eural Networks},
  year =	 2022,
  volume =	 33,
  number =	 4,
  pages =	 {1452--1466},
  doi =		 {10.1109/TNNLS.2020.3042395}
}

@inproceedings{rao2021msa,
  title =	 {MSA Transformer},
  author =	 {Rao, Roshan M and Liu, Jason and Verkuil, Robert and
                  Meier, Joshua and Canny, John and Abbeel, Pieter and
                  Sercu, Tom and Rives, Alexander},
  booktitle =	 {Proceedings of the 38th International Conference on
                  Machine Learning},
  pages =	 {8844--8856},
  year =	 2021,
  editor =	 {Meila, Marina and Zhang, Tong},
  volume =	 139,
  series =	 {Proceedings of Machine Learning Research},
  month =	 07,
  publisher =	 {PMLR}
}

@article{douglas2021phylodynamics,
  author =	 {Douglas, Jordan and Mendes, Fábio K and Bouckaert,
                  Remco and Xie, Dong and Jiménez-Silva, Cinthy L and
                  Swanepoel, Christiaan and de Ligt, Joep and Ren,
                  Xiaoyun and Storey, Matt and Hadfield, James and
                  Simpson, Colin R and Geoghegan, Jemma L and
                  Drummond, Alexei J and Welch, David},
  title =	 {{P}hylodynamics reveals the role of human travel and
                  contact tracing in controlling the first wave of
                  {COVID}-19 in four island nations},
  journal =	 {Virus Evolution},
  volume =	 7,
  number =	 2,
  pages =	 {veab052},
  year =	 2021,
  month =	 06,
  doi =		 {10.1093/ve/veab052}
}

@article{scheurer2025uncertainty,
  author =	 {{Scheurer}, Stefania and {Reiser}, Philipp and
                  {Br{\"u}nnette}, Tim and {Nowak}, Wolfgang and
                  {Guthke}, Anneli and {B{\"u}rkner}, Paul-Christian},
  title =	 {{U}ncertainty-{A}ware {S}urrogate-based {A}mortized
                  {B}ayesian {I}nference for {C}omputationally
                  {E}xpensive Models},
  journal =	 {arXiv e-prints},
  year =	 2025,
  month =	 may,
  doi =		 {10.48550/arXiv.2505.08683}
}

@article{penn2024phylo2vec,
  author =	 {Penn, Matthew J and Scheidwasser, Neil and Khurana,
                  Mark P and Duch\^{e}ne, David A and Donnelly,
                  Christl A and Bhatt, Samir},
  title =	 {{P}hylo2{V}ec: {A} {V}ector {R}epresentation for
                  {B}inary Trees},
  journal =	 {Systematic Biology},
  volume =	 74,
  number =	 2,
  pages =	 {250--266},
  year =	 2024,
  month =	 06,
  doi =		 {10.1093/sysbio/syae030}
}

@inproceedings{tagasovska2019single,
  title =	 {{S}ingle-model uncertainties for deep learning},
  author =	 {Tagasovska, Natasa and Lopez-Paz, David},
  booktitle =	 {Advances in Neural Information Processing Systems},
  volume =	 32,
  year =	 2019,
  doi =		 "10.48550/arXiv.1811.00908"
}

@article{voznica2022deep,
  author =       "Voznica, J. and Zhukova, A. and Boskova, V. and Saulnier, E.
                  and Lemoine, F. and Moslonka-Lefebvre, M. and Gascuel, O.",
  title =        {{D}eep learning from phylogenies to uncover the
                  epidemiological dynamics of outbreaks},
  journal =      "Nature Communications",
  year =         2022,
  month =        07,
  day =          06,
  volume =       13,
  number =       1,
  pages =        3896,
  doi =          "10.1038/s41467-022-31511-0"
}

@article{lambert2023deep,
  author =	 {Lambert, Sophia and Voznica, Jakub and Morlon,
                  H\'{e}l\`{e}ne},
  title =	 {{D}eep {L}earning from {P}hylogenies for
                  {D}iversification {A}nalyses},
  journal =	 {Systematic Biology},
  volume =	 72,
  number =	 6,
  pages =	 {1262--1279},
  year =	 2023,
  month =	 08,
  doi =		 {10.1093/sysbio/syad044}
}

@article{frasconi1998general,
  author =       {Frasconi, P. and Gori, M. and Sperduti, A.},
  journal =      {IEEE Transactions on Neural Networks},
  title =        {{A} general framework for adaptive processing of data
                  structures},
  year =         1998,
  volume =       9,
  number =       5,
  pages =        {768--786},
  doi =          {10.1109/72.712151}
}

@article{lajaaiti2023comparison,
  author =	 {Isma{\"e}l Lajaaiti and Sophia Lambert and Jakub
                  Voznica and H{\'e}l{\`e}ne Morlon and Florian
                  Hartig},
  title =	 {{A} {C}omparison of {D}eep {L}earning
                  {A}rchitectures for {I}nferring {P}arameters of
                  {D}iversification {M}odels from {E}xtant
                  Phylogenies},
  year =	 2023,
  doi =		 {10.1101/2023.03.03.530992},
  publisher =	 {Cold Spring Harbor Laboratory},
  journal =	 {bioRxiv}
}

@article{leroy2025graph,
  author =	 {Leroy, Am{\'e}lie and Lajaaiti, Isma{\"e}l and
                  Lambert, Sophia and Voznica, Jakub and Pichler,
                  Maximilian and Morlon, H{\'e}l{\`e}ne and Hartig,
                  Florian and Jacob, Laurent},
  title =	 {{G}raph {N}eural {N}etworks for {L}ikelihood-{F}ree
                  {I}nference in {D}iversification Models},
  year =	 2025,
  doi =		 {10.1101/2025.08.14.670341},
  publisher =	 {Cold Spring Harbor Laboratory},
  journal =	 {bioRxiv}
}

@book{stadler2024decoding,
  author =	 {Tanja Stadler and Carsten Magnus and Timothy Vaughan
                  and Jo{\"e}lle Barido-Sottani and Veronika
                  Bo{\v{s}}kov{\'a} and Jana S. Huisman and J{\=u}lija
                  Pe{\v{c}}erska},
  title =	 {{D}ecoding {G}enomes: {F}rom {S}equences to
                  {P}hylodynamics},
  year =	 2024,
  month =	 04,
  publisher =	 {ETH Zurich},
  doi =		 {10.3929/ethz-b-000664449},
  url =		 {https://decodinggenomes.org/}
}

@article{stadler2013birth,
  author =       {Stadler, Tanja and K{\"u}hnert, Denise and Bonhoeffer,
                  Sebastian and Drummond, Alexei J.},
  title =        {{B}irth-death skyline plot reveals temporal changes of
                  epidemic spread in {HIV} and hepatitis {C} virus ({HCV})},
  volume =       110,
  number =       1,
  pages =        {228--233},
  year =         2013,
  doi =          {10.1073/pnas.1207965110},
  publisher =    {National Academy of Sciences},
  journal =      {Proceedings of the National Academy of Sciences}
}

@article{gelman2020bayesian,
  author =	 {{Gelman}, Andrew and {Vehtari}, Aki and {Simpson},
                  Daniel and {Margossian}, Charles C. and {Carpenter},
                  Bob and {Yao}, Yuling and {Kennedy}, Lauren and
                  {Gabry}, Jonah and {B{\"u}rkner}, Paul-Christian and
                  {Modr{\'a}k}, Martin},
  title =	 {{B}ayesian {W}orkflow},
  journal =	 {arXiv e-prints},
  year =	 2020,
  month =	 nov,
  doi =		 {10.48550/arXiv.2011.01808},
}

@article{zammitmangion2025neural,
  author =	 {Zammit-Mangion, Andrew and Sainsbury-Dale, Matthew
                  and Huser, Rapha\"{e}l},
  title =	 {{N}eural {M}ethods for {A}mortized {I}nference},
  journal =	 "Annual Review of Statistics and Its Application",
  year =	 2025,
  volume =	 12,
  pages =	 "311--335",
  doi =		 "10.1146/annurev-statistics-112723-034123",
  publisher =	 "Annual Reviews",
}

@article{sainsburydale2023likelihood,
  author =	 {Matthew Sainsbury-Dale and Andrew Zammit-Mangion and
                  Rapha\"{e}l Huser},
  title =	 {{L}ikelihood-{F}ree {P}arameter {E}stimation with
                  {N}eural {B}ayes {E}stimators},
  journal =	 {The American Statistician},
  volume =	 78,
  number =	 1,
  pages =	 {1--14},
  year =	 2024,
  publisher =	 {ASA Website},
  doi =		 {10.1080/00031305.2023.2249522}
}

\newpage

\section{Supplementary Information}

\subsection{Simulation}\label{sec:si:simulation}

Each simulation in our database is obtained by sampling parameters
from a prior distribution and then sampling from a
birth-death-sampling process with those parameters. The details of
this process are enumerated below, and we summarize the distributions used in
\SITabref{tab:si:prior-reduced}. The prior distribution on the model parameters is
adapted from an analysis of SARS-CoV-2 genomes in
\cite{douglas2021phylodynamics}.

\begin{enumerate}
  \item Sample a stopping time for the simulation of 30--90 days:
    \[\stoppingTime\sim\text{Discrete-uniform}\{30,\ldots,90\}.\]
  \item Sample a constant net becoming-uninfectious rate
    from a log-normal distribution:
    \(\sigma\sim\Lognormal{-1.81}{0.2}\). This prior has been scaled
    from the one used in \cite{douglas2021phylodynamics} to be in
    units of days (rather than years).
  \item Sample random change points for the piece-wise constant functions by first drawing a
    number of change points from a discrete-uniform distribution on \({1,2}\), and then sampling
    that number of change times from the
    distribution \(\Uniform{0}{\stoppingTime}\).
  \item Sample the values for the piece-wise constant function describing the reproduction number, \(\reff\sim\Lognormal{1.0}{0.7}\)
  \item Sample the values of the piece-wise constant proportion of infections that are observed, \(p_{\psi}\sim\Beta{1.1}{8.0}\). It is this part of the process that is changed in the alternative simulation described in \SISecref{sec:si:alt-model}.
  \item Compute the birth, death and sampling rates, \(\lambda\), \(\mu\) and
    \(\psi\) as piece-wise constant functions from the
    sampled values of \(\reff\), \(\sigma\) and \(p_{\psi}\).
  \item Sample from the birth-death-sampling process, which can be described with the
    following equations:
    \begin{description}
      \item [Birth] \(X \overset{\lambda}{\longrightarrow} 2X\)
      \item [Death] \(X \overset{\mu}{\longrightarrow} \emptyset\)
      \item [Sampling] \(X \overset{\psi}{\longrightarrow} \text{Sequence}\)
    \end{description}
    This process continues until one of three conditions are met: the
    stopping time, \(\stoppingTime\) is reached, the prevalence of infection reaches
    50,000, or there are 1,000 sequenced infections. The realizations
    of this process are conditioned to have at least two sequenced
    infections and for the pathogen not to have gone extinct before
    the end of the process.
  \item Compute the reconstructed tree by tracking transmission among the individuals in the birth-death-sampling
    process and who was sampled. We can reconstruct a transmission tree
    (\Figref{fig:timelines}A) describing who-infected-whom and the
    circumstances under which they ceased to be infectious (i.e.\ whether they were sampled when they ceased to be infectious). This can then be pruned to obtain a reconstructed tree
    connecting the sequenced infections (\Figref{fig:timelines}B).
  \item Generate a collection of \emph{measurements}
    of the simulation. Each measurement is a vector
    \[M_{i}=(t_m,\numPrev{t_m},\numCum{t_m},\refft{t_m})\] where we
    have sampled the measurement time randomly \(t_m\sim\Uniform{0}{t_{\text{present}}}\) (see
    \Figref{fig:timelines}.)
  \item Record the set of measurements, \(\{M_{i}^{(j)}\}\) and the corresponding
    reconstructed tree, \(\tree^{(j)}\), from this \(j\)th realization
    of the process. By repeating this \(\numTrain\) many times, we
    generate our training dataset:
    \(\dataTraining=\{(\tree_j,\{M_{i}^{(j)}\})\}\).
\end{enumerate}

\begin{table}[ht]
  \centering
  \begin{tabular}{llll}
    \hline
    \textbf{Quantity}           & \textbf{Prior}                   & \begin{tabular}[c]{@{}l@{}}\textbf{Median}\\\textbf{(95\% interval)}\end{tabular} \\ \hline
    \textbf{Model parameters}   &                                  &                                                                    \\
    Reproduction number         & \(\Lognormal{1.0}{0.7}\)         & 2.72 (0.69, 10.72)                                                       \\
    Net removal rate            & \(\Lognormal{-1.81}{0.2}\)       & 0.16 (0.11, 0.24)                                                 \\
    Sampled proportion          & \(\Beta{1.1}{8.0}\)              & 0.094 (0.005, 0.384)                                                        \\
    \textbf{Hyperparameters}    &                                  &                                                                    \\
    \begin{tabular}[c]{@{}l@{}}Epidemic duration\\(in days)\end{tabular} & \(\text{Discrete-uniform}\{30,\ldots,90\}\)             &                                                                    \\
    \begin{tabular}[c]{@{}l@{}}Number of parameter\\changes\end{tabular} & \(\text{Discrete-uniform}\{1,2\}\)               &                                                                    \\
    Change times                & \(\Uniform{0}{T_{\text{stop}}}\) &                                                                    \\ \hline
  \end{tabular}
  \caption{\label{tab:si:prior-reduced} The prior distribution used in the simulations along with summary statistics to indicate typical samples.}
\end{table}

\subsection{Sensitivity analysis}

\subsubsection{Alternative model with delayed sampling}\label{sec:si:alt-model}

As a sensitivity analysis, we consider an alternative simulation model in which there is \emph{delayed
sampling}: there is an initial interval during which the sampling rate is
set to zero, after which sampling is activated and remains constant. The sampled proportion (i.e., the sampling intensity
once activated) is drawn from the same marginal distribution used to generate the
sampling proportions for the original training dataset.

The activation time of the sampling is chosen uniformly as a proportion of the full
epidemic duration, with
the proportion drawn from $t_{\text{act}}\sim\Uniform{0.3}{0.7}$. Consequently, this
alternative model generates datasets in which there is a period of
unobserved transmission early in the epidemic, during which infections
accumulate but no sequences are observed.

All other aspects of the simulation model are defined as in the original
simulation: the reproduction number, $\reff$, is specified in the same way
(piece-wise constant between change times, with values drawn from the
prior), and the net removal rate is once again constant with
$\sigma \sim \Lognormal{-1.81}{0.2}$. The key difference is that the
sampling proportion is a piece-wise function with a single change point: before
the activation time, $t<t_{\text{act}}T_{\text{stop}}$, the sampling proportion is $p_{\psi}(t)=0$, and after activation, $t\geq t_{\text{act}}T_{\text{stop}}$,
it is constant, $p_{\psi}(t)=p_{\psi}^{i}$, with $p_{\psi}^{i}\sim\Beta{1.1}{8.0}$.

\subsubsection{Sensitivity to choice of prior distribution}\label{sec:si:prior-sensitivity}

\begin{figure}[h!]
  \centering
  \begin{subfigure}{0.44\textwidth}
    \includegraphics[width=\linewidth]{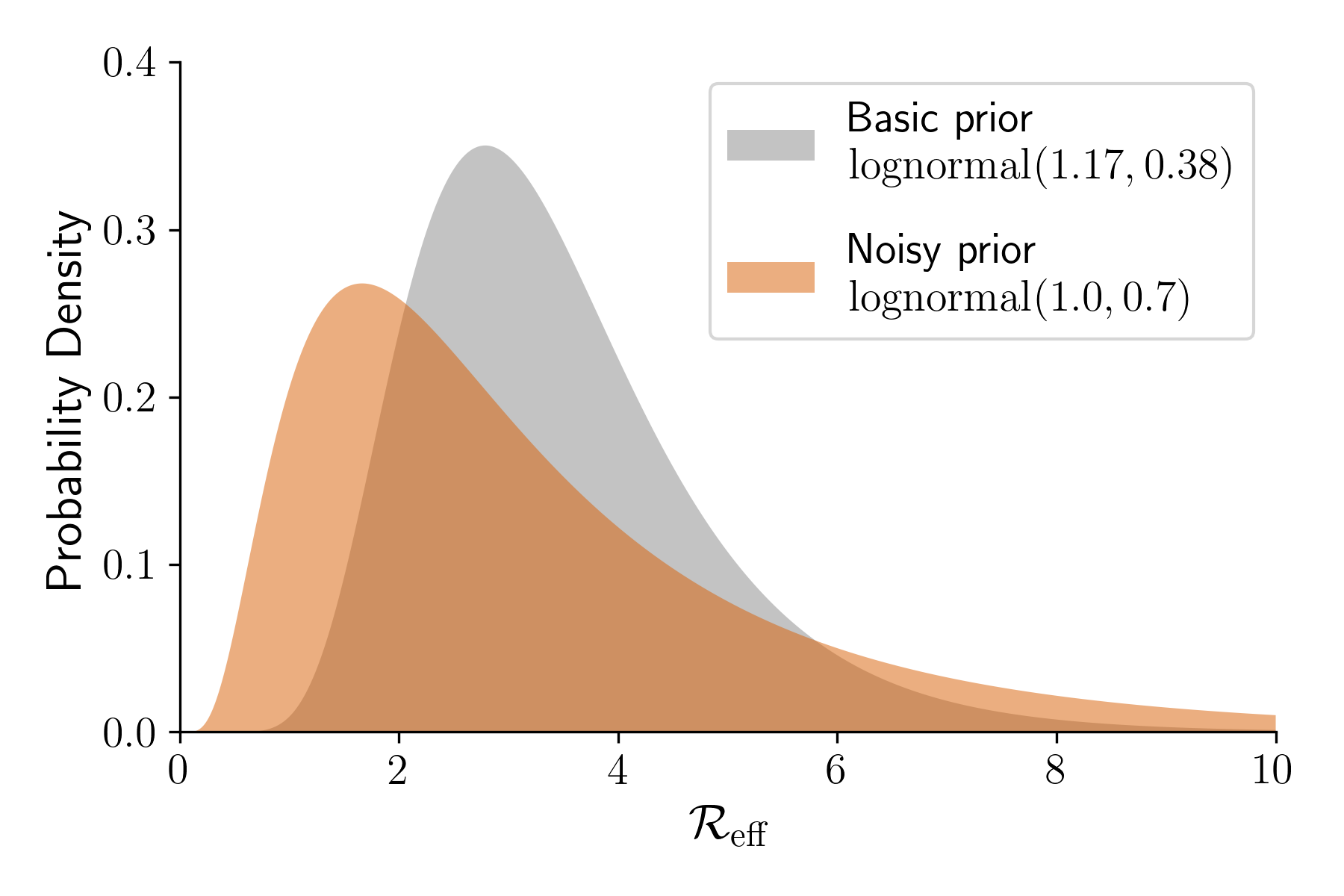}
    \caption{$\reff$ prior distributions}
  \end{subfigure}
  \hspace{1cm}
  \begin{subfigure}{0.44\textwidth}
    \includegraphics[width=\linewidth]{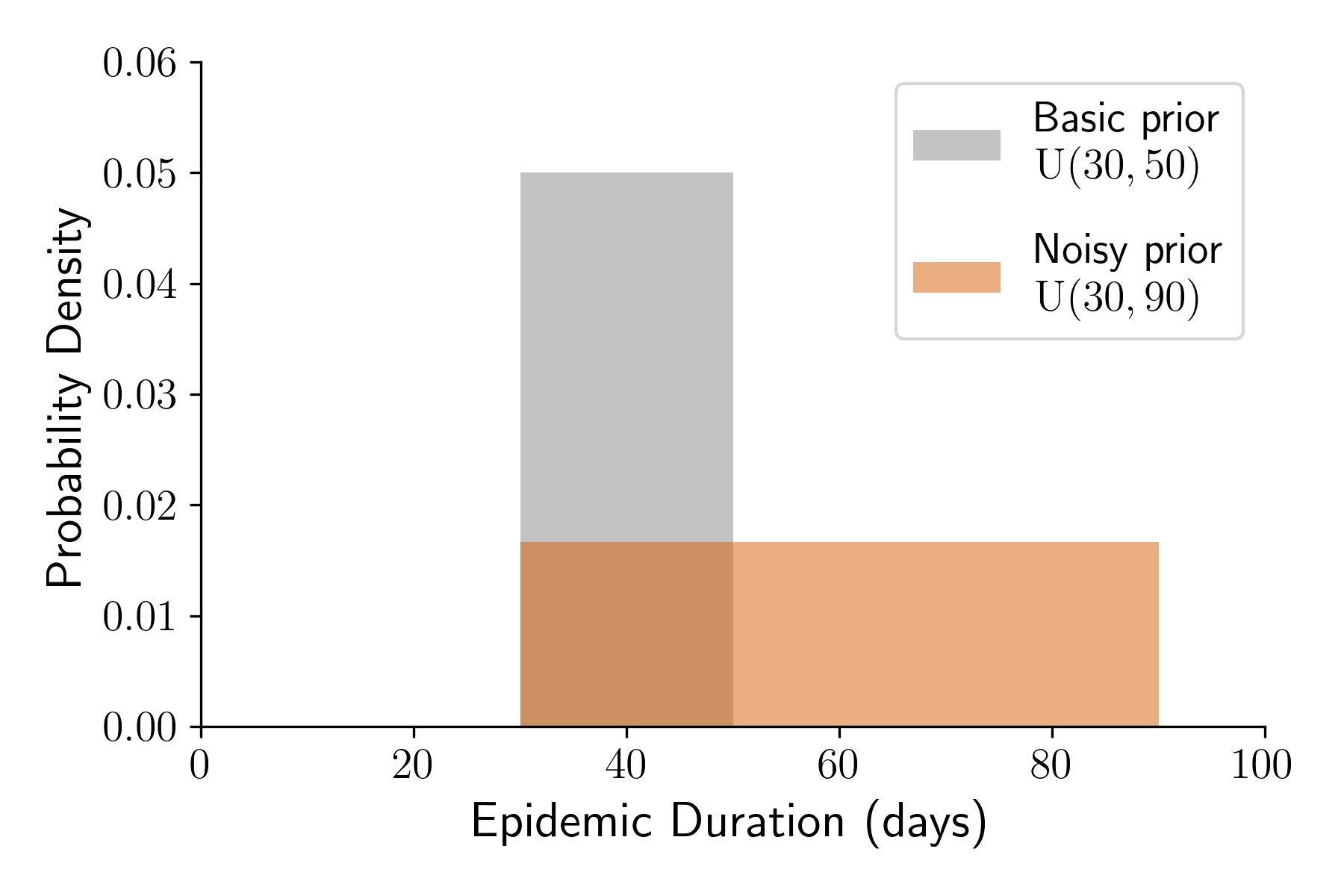}
    \caption{Epidemic duration prior distributions}
  \end{subfigure}
  \begin{subfigure}{0.9\textwidth}
    \includegraphics[width=\linewidth]{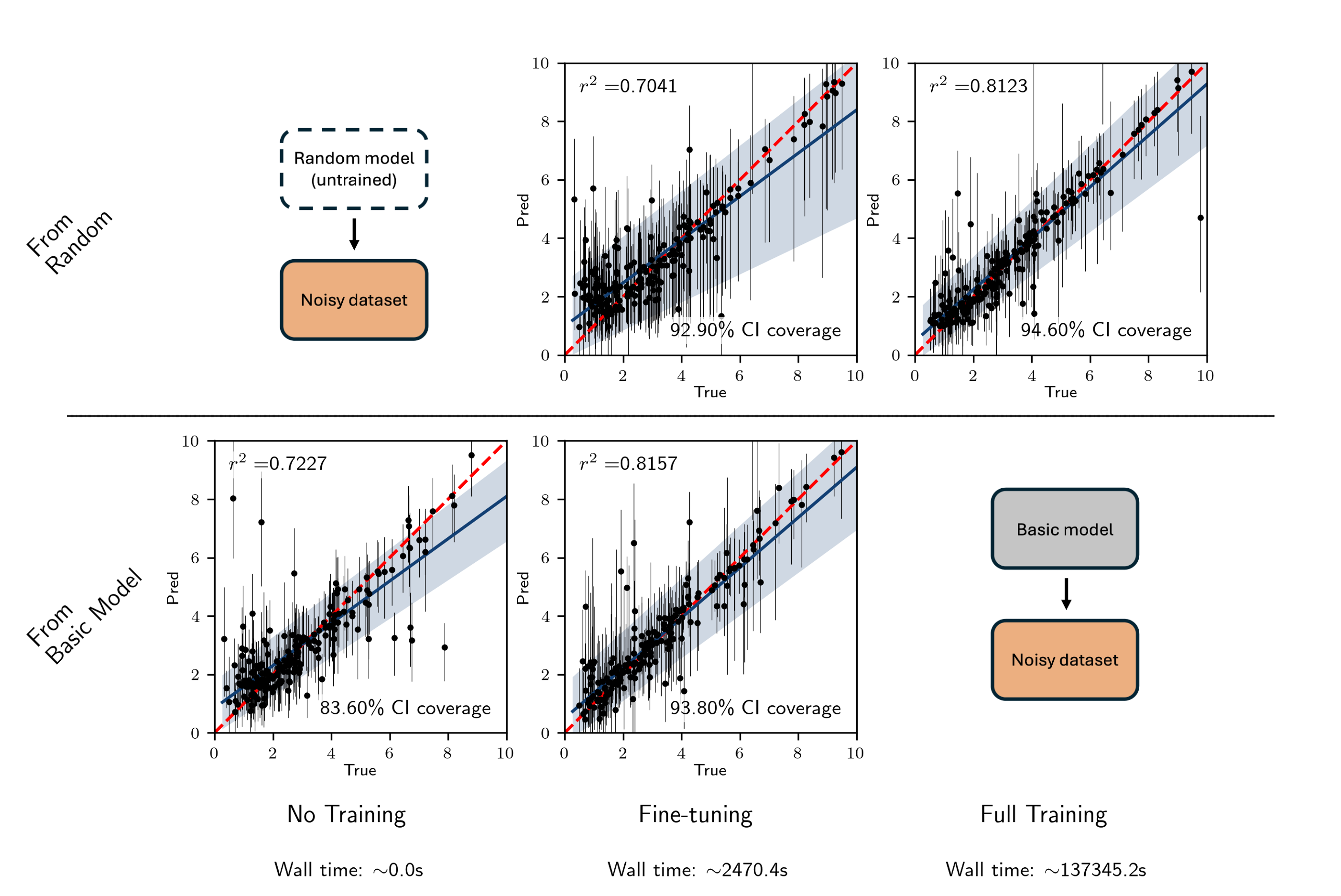}
    \caption{Estimates from model variants}
  \end{subfigure}
  \caption{\label{fig:fine_tuning_r0} NBEs can be rapidly
    retrained to adapt to different prior distributions.
    \textbf{(a)} and \textbf{(b)} show the basic and noisy
    distributions considered. \textbf{(c)} shows the true
    values and the estimates of the basic reproduction number
    sampled from the noisy prior. The models used to generate
    the estimates start from either a random initialization (top
    row) or a model pre-trained on data sampled from the basic
    prior. We show the performance for three levels of training:
    no training (for the basic model), fine-tuning (for both
    models), and full training (from the random model). For each
    level of training we also show the indicative wall time
  required.}
\end{figure}

A further simulation study was carried out to assess the impact of
misspecifying the prior distribution. We considered two potential
prior distributions: a \emph{basic} distribution, and a more diffuse
\emph{noisy} distribution. In \SIFigref{fig:fine_tuning_r0} we show the
basic and noisy prior distributions for $\reff$ and the epidemic
duration, $T_{\text{stop}}$. Prior distributions for all other
parameters are the same between the two cases. We considered four
variations of the NBE in this sensitivity analysis:

\begin{description}
  \item[Zero-shot pre-trained] a model trained on the basic
    distribution which is then applied to data drawn from the noisy
    distribution;
  \item[Fine-tuned pre-trained] a model trained on the basic
    distribution and subsequently fine-tuned on data from the noisy distribution is then applied to data from the noisy distribution;
  \item[Fine-tuned from random] a model in which the weights are
    randomly initialized, and only the weights of \(\fpred\) are
    modified during training is then applied to data simulated from the noisy distribution; and
  \item[Full training from random] a model in which the weights are
    randomly initialized and the model is trained on data simulated
    from the noisy distribution, which is then applied to data simulated from the noisy distribution.
\end{description}
\noindent
In \SITabref{tab:sa-priors-summary} we present the time taken to
pre-train or fine-tune each model variant and the resulting
performance as measured by the coefficient of determination, $R^2$,
for the estimates of $\refft{t}$, $\logten{\numPrev{t}}$, and
$\logten{\numCum{t}}$. This demonstrates that fine-tuning an existing
model takes orders of magnitude less time than training a model from
scratch and results in comparable performance.

\begin{table}[h]
  \begin{tabular}{lllll}
    \hline
    \textbf{Training}  & \textbf{
      \begin{tabular}[c]{@{}l@{}}Wall time\\ (seconds)
    \end{tabular}} & \textbf{
      \begin{tabular}[c]{@{}l@{}}\(\reff\)\\($R^2$)
    \end{tabular}} & \textbf{
      \begin{tabular}[c]{@{}l@{}}\(\logten{\numPrevX}\)\\($R^2$)
    \end{tabular}} & \textbf{
      \begin{tabular}[c]{@{}l@{}}\(\logten{\numCumX}\)\\($R^2$)
    \end{tabular}} \\ \hline
    \begin{tabular}[c]{@{}l@{}}Zero-shot\\pre-trained\end{tabular}         & \textbf{0.0}          & 0.723              & 0.924             & 0.933                \\
    \begin{tabular}[c]{@{}l@{}}Fine-tuned\\pre-trained\end{tabular}        & \underline{2315.6}    & \textbf{0.816}     & \underline{0.944} & \underline{0.952}    \\
    \begin{tabular}[c]{@{}l@{}}Fine-tuned\\from random\end{tabular}        & 2470.4                & 0.704              & 0.817             & 0.860                \\
    \begin{tabular}[c]{@{}l@{}}Full training\\from random\end{tabular}     & 137345.2              & \underline{0.812}  & \textbf{0.955}    & \textbf{0.962}       \\ \hline
  \end{tabular}
  \caption{\label{tab:sa-priors-summary} The NBE performs
    reasonably well when applied to test data drawn from a more diffuse
    prior (as depicted in \SIFigref{fig:fine_tuning_r0}). For each model, this shows the training time and the coefficient of determination, $R^2$ for each of the quantities of interest: \(\reff\), the reproduction number; \(\logten{\numPrevX}\), the common logarithm of the prevalence; and \(\logten{\numCumX}\), the common logarithm of the cumulative incidence. With a small
    amount of fine-tuning of the neural network, the estimators
    performance matches that of a model trained from scratch, suggesting
  a way forward for re-use of NBEs.}
\end{table}

\FloatBarrier

\subsection{BEAST2 MCMC using BDSky}\label{sec:si:bdsky}

For the MCMC baseline, we used the Birth-Death Skyline (BDSky) tree prior \citep{stadler2013birth} implemented in BEAST2 \citep{bouckaert2019beast}. To align this analysis with the
simulation study and isolate inference of epidemiological parameters, we fixed
the phylogeny to the true reconstructed tree and fixed the becoming-uninfectious
rate at its true value. The MCMC therefore estimated only two quantities: the
effective reproduction number and the sampling proportion. The effective
reproduction number was parameterized as a piece-wise constant trajectory with
10 values uniformly spaced through time, while the sampling proportion was
modeled as a single constant parameter over time.

For each of the testing datasets, we ran one chain for 5,000,000 iterations and
retained 5,000 posterior samples via thinning. For all runs, effective sample
sizes were greater than 200. Priors for the values of both the reproduction number and the
sampling proportion matched those used when generating the testing data (\SITabref{tab:si:prior-reduced}).

\end{document}